%% file: ITW_arxiv-v2.tex
\documentclass[11pt,a4paper,draftclsnofoot,onecolumn]{IEEEtran}

	\input{settings_preamble}

	\usepackage{setspace}
	\input{./Definitions}

	\usepackage{microtype}

	\usepackage[pdfencoding=auto,psdextra,colorlinks=true,linkcolor = blued,citecolor = blued,runcolor = cyan,urlcolor = cyan!90!green!30!black]{hyperref}

	\newcommand{\sdk}{{_{d_k}}}
	\newcommand{\Zk}{{\Zc_{[k]}}}
	\newcommand{\Wk}{{\Wc_{[k]}}}
	\newcommand{\Wkm}{{\Wc_{[k-1]}}}
	\newcommand{\minkn}{{K'}}
	\newcommand{\dec}{{\text{dec}}}
	\newcommand{\ang}[1]{{\ensuremath{[#1]}}}
	\DeclareMathOperator{\NDT}{{NDT}}
	
\IEEEoverridecommandlockouts

\begin{document}
 \title{Rate-Memory Trade-Off for the Cache-Aided MISO Broadcast Channel with Hybrid CSIT}

\author{\thanks{This work is supported by the European Research Council under the EU Horizon 2020 research and innovation program / ERC grant agreement no. 725929 (ERC project DUALITY).}
\IEEEauthorblockN{Antonio Bazco-Nogueras, Petros Elia}\\
\IEEEauthorblockA{Communication Systems Department, \
EURECOM \\ 
Email: \{antonio.bazco-nogueras; \ petros.elia\}@eurecom.fr}}

	\maketitle

	\begin{abstract} 
		One of the famous problems in communications was the so-called ``PN'' problem in the Broadcast Channel, which refers to the setting where a fixed set of users provide perfect Channel State Information (CSI) to a multi-antenna transmitter, whereas the remaining users only provide finite precision CSI or no CSI. 
		The Degrees-of-Freedom (DoF) of that setting were recently derived by means of the Aligned Image Set approach.
		In this work, we resolve the cache-aided variant of this problem (i.e., the ``PN'' setting with side information) in the regime where the number of users providing perfect CSI is smaller or equal to the number of transmit antennas. 
		In particular, 
		we derive the optimal rate-memory trade-off under the assumption of uncoded placement, and characterize the same trade-off within a factor of~2.01 for general placement. 
		The result proves that the ``PN'' impact remains similar even in the presence of side information, but also that the optimal trade-off is not achievable through independently serving the two sets of users. 			
	\end{abstract}

		\section{Introduction}\label{se:intro}
		Coded caching has emerged as a promising tool for coping with the challenging increase of content demand in wireless networks. 
		Initially, the benefit of coded caching was identified for the setting in which a server communicates to $K$ users through an error-free single-stream shared link of fixed capacity\cite{MaddahAli2014,Yu2018_TIT,Yu2019_TIT}. 			
		In this setting, the server has access to a library of $N$ files, and each user has access to a local memory (cache) of size equal to the size of $M$ files, where it can store content from the library. 
		For the aforementioned scenario, it was shown in\cite{MaddahAli2014,Yu2018_TIT,Yu2019_TIT} that coded caching provides a speed-up factor (or \emph{coded-caching gain}) of $K\frac{M}{N}+1$ as compared to the uncoded caching case, since it allows us to simultaneously serve $K\frac{M}{N}+1$ users.   
		
		The promising gains of coded caching in this initial setting fostered the interest in understanding how these gains could be translated into wireless networks, which was analyzed e.g. by considering 
		uneven link capacities\cite{Joudeh2020_mixed,Piovano2019} or multi-antenna transmissions\cite{Naderializadeh2017,lampiris2020resolving}. 
		The context of multi-antenna coded caching has recently received considerable attention, especially after the latest results in~\cite{Lampiris2018_JSAC}, which reveal that multi-antenna coded caching systems provide all the massive gains of coded caching, together with all the multiplexing gains of multi-antenna systems, and can do so without the hindrance of the 
		subpacketization bottleneck that has previously kept such technologies from taking off. 		
	
		
		In such multi-antenna context, it is crucial to 
		characterize the impact of CSI availability. 
		Toward this, among other works, \cite{Piovano2019} considered partial CSI at the transmitter (CSIT) on the cache-aided MISO Broadcast Channel (BC),  \cite{lampiris2020resolving} analyzed the CSIT required to maintain the ideal caching gains in multi-antenna settings, \cite{lampiris2017_ISIT_noCSIT} studied cache-aided interference management with no CSIT, and \cite{Zhang2017} focused on the delayed CSIT setting. 
	
		In this work, we consider a cache-aided variant of a classical CSI-related problem~in multi-antenna communications: 
		the cache-aided MISO BC~in which the $L$-antenna transmitter has perfect CSI only for~the channel of a \emph{fixed} set of $K_P$ users, whereas the CSI of the other $K-K_P$ users is available only up to finite precision\footnote{We define  the finite precision channel knowledge in Def.~\ref{def:bounded} as being equivalent to assume that, for the transmitter, all joint and conditional probability density functions of channel coefficients exist and are bounded.
		We note that the DoF results are analogous for finite precision CSI and for no CSI\cite{Davoodi2016_TIT_DoF}. Thus, we omit the latter for  concision.}~\cite{Davoodi2016_TIT_DoF}. 
		This scenario is equivalent to the so-called  ``PN'' BC setting\cite{Tandon2012b,Lashgari2016,Davoodi2016_TIT_DoF}, 
		but with the non-trivial extension of considering side information at the receivers. 
		The original  ``PN'' BC setting without side information  remained an open problem for many years, and its DoF were finally derived by Davoodi and Jafar in~\cite{Davoodi2016_TIT_DoF} by means of the Aligned Image Set (AIS) approach.  
		By incorporating the AIS approach into the derivation of the delivery time of coded caching, we obtain the optimal rate-memory trade-off of this setting under the assumption of uncoded placement when $K_P\leq L$, and characterize the same trade-off within a factor of~$2.01$ for~general placement. 
		Moreover, we show that the optimal trade-off is not~achievable through separated transmission to the two sets of users. 
				
			\subsubsection*{Notation}\label{se:notation}
				We make use of the Landau notation, and consequently, for any two functions $f(x),\; g(x)$, the expression $f(x) = o(g(x))$ stands for $\lim_{x\to 0}\frac{f(x)}{g(x)} = 0$.
				For any natural number $n$, $\ang{n}$ stands for $\ang{n} = \{1,\, 2,\, \dots,\, n\}$. 


		\section{System Model and Problem Definition}\label{se:sys}

			\subsection{Communication Channel} \label{se:trans-sys}
			We consider the  $L\times K$ MISO BC in which a transmitter (TX) with $L$ transmit antennas serves $K$ single-antenna cache-aided users. The received signal at user~$i$ is written as
				\eqm{
						Y_i(t) = \sum\nolimits_{j=1}^Lh\sij(t) X_j(t) + \zeta_i(t),
				}
			where $X_j(t)$ is the transmit signal from the $j$-th antenna of the transmitter, normalized such that $\Exp[|X_j(t)|^2]\leq P$, and where $P$ is the nominal SNR value which is allowed in the DoF framework to approach infinity\cite{Davoodi2016_TIT_DoF}. 
			Furthermore, $\zeta_i(t)$ is the i.i.d. additive white Gaussian noise (AWGN), and  
			$h\sij$ denotes the fading coefficient between the $j$-th antenna of the TX and  user~$i$. 
			We assume that $|h\sij|$ is bounded away from zero and infinity. 
			The channel vector from the $L$ transmit antennas to user~$i$ is denoted by $\hv_i$, and the global channel matrix by $\bH$. 			
			We further define the vectors $\bX(t) \triangleq (X_1(t),\,\dots,\,X_L(t))$, $X_j^{[\tau]} \triangleq \big\{X_j(t)\big\}_{t\in[\tau]}$,   $\bX^{[\tau]} \triangleq \big\{\bX(t)\big\}_{t\in[\tau]}$, and $Y_i^{[\tau]} \triangleq \big\{Y_i(t)\big\}_{t\in[\tau]}$.

			\subsection{Cached-aided Network}\label{se:caching}
			We consider a cached-aided scenario in which the TX has access to a content library of $N$ files, each one of size equal to $B$ bits. We assume that $N\geq K$, and  
			we denote the $n$-th file of the library as $W_n$, $n\in[N]$. 
			Each user has a local memory cache of size $MB$ bits, where $M\in\{0,1,\dots,N\}$, 
			in which it stores (coded or uncoded) data from the library. 
		  We denote the normalized cache size with respect to the library size as $\gamma \triangleq \frac{M}{N}$, and the cache available at  user~$i$ as $Z_i$, such that $|Z_i|=MB$.
			
			Coded caching systems operate in two  phases: the \emph{placement phase},	in which the users fill their cache with content from the library, and the \emph{delivery phase}, 
			in which each user requests a  library file to be delivered by the TX.  
			The file requested by user~$k$ is denoted as $W_{d_k}$, $d_k\in[N]$, and the vector of requested file indexes  
			is denoted by $\bd\triangleq (d_1,\dots,d_K)\in[N]^K$. 
			We allow coded cache placement, but we also analyze the particular case in which only uncoded placement is allowed.\footnote{A prefetching scheme is called an \emph{uncoded prefetching scheme} if each user stores $MB$ bits from the database without coding.}

			\subsection{Hybrid CSIT}\label{se:pn}
			We focus on the $K$-user MISO BC scenario in which the TX only obtains perfect CSI for the channel of a subset of $K_P$ users, where $K_P\leq L$.\footnote{Cases where $K_P=0$ or $K_F =0$ are not comprised within the cases with \emph{hybrid} CSIT, but we include them for completeness.} In turn, the CSI of the other $K_F \triangleq K-K_P$ users is available only up to finite precision. 
			Finite precision CSIT refers to the case in which, from the transmitter's perspective, the channel coefficients satisfy the bounded density assumption which is defined as follows. 
			
			\begin{definition}[\!\!{\cite[Definition 4]{Davoodi2020_sumset}}~Bounded Density Channel Coefficients]\label{def:bounded}
					The set of random variables~$\Hc$ satisfies the bounded density assumption if  both of the following conditions hold.
					\enb
						\item The magnitudes of all the random variables in $\Hc$ are bounded away from zero and infinity, i.e., there exist two constants $\Delta_1,\,\Delta_2$ such that for all $h \in \Hc$ it holds that  $0<\Delta_1 < |h|\leq \Delta_2<\infty$.  
						\item There exists a finite positive constant $f_{\max}$, such that for all finite cardinality disjoint subsets $\Hc_1, \Hc_2$ of $\Hc$, the joint probability density function of all random variables in $\Hc_1$, conditioned on all random variables in $\Hc_2$, exists and is bounded above by $f_{\max}^{|\Hc_1|}$.
					\ene
			\end{definition}			
		
		Although~\cite{Piovano2019} studied also the cache-aided MISO BC, in~\cite{Piovano2019} all users provide the same level of (possibly imperfect) CSI, while here we consider different CSIT level for different users. 
		We remark that this setting is also fundamentally different from the one considered in~\cite{lampiris2020resolving}, 
			where it was shown that, in the $L\times K$ cache-aided MISO BC, we can achieve the optimal coded-caching gain under one-shot linear schemes ($K\gamma+L$) even if the TX  has CSI for only $L$ served users \emph{at a time}. 
			The difference resides in the fact that, in~\cite{lampiris2020resolving}, the TX needs CSI from only $L$ users at a time, but the users who are providing the CSI can (and do) change when the subset of served users changes. Hence,~\cite{lampiris2020resolving} derives the minimum instantaneous CSI requirements for a setting where all users must eventually provide CSI. 
			In contrast, we consider in this work that a \emph{fixed} subset of users provide CSI, and this subset does not change throughout the whole transmission. Thus, there are two classes of users according to the quality of the provided CSI, which is a setting that models scenarios with heterogeneous devices where some of these devices have reduced capabilities or operate under simpler~protocols. 


			\subsection{Problem Definition}\label{se:prob}
			We present  the formal definition of the rate-memory trade-off considered. 
			Let $\{W_n\}_{n\in[N]}$ be $N$ i.i.d. random variables, each uniformly distributed over $[2^{\LF B\RF}]$, and let us introduce the notation $\Wk = \{W_n\}_{n\in[k]}$, $k\leq N$. 
			We scale the size of the files proportionally to the number of channel uses $\tau$ as $B\triangleq \tau R$, where $R$ is the transmission rate in bits per channel~use. 
			
			A code $\Cc(\tau, R, M)$  consists of a prefetching strategy, an encoding scheme and $K$ decoding functions, which are explained in the following.  
			We characterize a prefetching strategy $\bphi$ by its $K$ caching functions $\phi_k$, $\bphi\triangleq (\phi_1, \dots, \phi_K)$, each of which is a function that maps the library content into the cache content of one user during the placement phase. 
			Thus,
					$\phi_k : [2^{\LF B\RF}]^N \rightarrow [2^{\LF MB\RF}]$, and $Z_k \triangleq  \phi_k(W_1,\, \dots,\,W_N)$. 
			The encoding scheme $\psi : [N]^K \times [2^{\LF B \RF}]^N\rightarrow \Cb^{L\times \tau}$ maps a demand vector $\bd$ and $N$ files into a codeword vector 
					$\bX^{[\tau]} \triangleq  \psi(\bd, \Wc_{[N]})$
			satisfying the average power constraint~$\Exp[|X_j(t)|^2]\leq P$. 
			Finally, a decoding function $\mu_{k} : [N]^K \times \Cb^{\tau} \times  [2^{\LF MB\RF}] \rightarrow [2^{\LF B\RF}]$ maps a requested demand, the signal received $Y_k^{[\tau]}$, and the cache content $Z_k$ into an estimate $\hW_{d_k} \triangleq \mu_{k}(\bd,Y_k^{[\tau]},Z_k)$ of $W_{d_k}$.  
			The probability of error is defined as \vspace{-.5ex}
				\eqm{ \label{eq:prob_error}
					P_{e,\tau} \triangleq \max_{\bd\in[N]^K} \max_{k\in[K]} \Pr(\hW_{d_k} \neq W_{d_k}).
				}
			Note that 
			\eqref{eq:prob_error} reflects a \emph{worst-case} metric over all possible demands. 
			Thus, we consider that each user requests a distinct~file. 
			
			We want to characterize the channel uses required to transmit a single bit of content to each user\cite{MaddahAli2014,Piovano2019}. 
			This delay will be referred to as the \emph{delivery time}, and it is given by $T \triangleq \frac{1}{R}$. 
			%
			More rigorously, for a given prefetching strategy $\bphi$,
			a delivery time $T_{\varepsilon,\phi}(M)$ is said to be $\varepsilon$-achievable if and only if, for every $\varepsilon > 0$ and big enough file size $B$, there exists a code $\Cc(\tau,R,M)$ with probability of error less than $\varepsilon$. 
			
			For a given memory constraint $M$, the rate-memory trade-off $T_{}^\star(M)$ is defined as the minimum delivery time  that can be achieved by \emph{any} prefetching scheme with vanishing error probability and for sufficiently large file size.
			Thus, it is defined as 
				\eqm{
					T_{}^\star(M) \triangleq \sup_{\varepsilon>0} \limsup_{B\rightarrow\infty}\min_{\phi} T^\star_{\epsilon,\phi}(M). \label{eq:def_tstar}
				}
			We omit hereinafter the dependence on $M$, unless it is necessary for clarity.
			
			In pursuit of highlighting the impact of the multi-user interference, 
			we consider the fundamental limit where  $P\rightarrow\infty$. 
			Therefore, we consider the optimal \emph{Normalized Delivery Time} (NDT), which is defined as the  ratio of the optimal delivery time $T^\star$ over the time required to deliver a single bit to a cache-less user, in the absence of interference, as $P$ approaches infinity\cite{Joudeh2020_mixed,Xu2017_TIT}.  
			Consequently, it is defined as 
				\eqm{ 
					\NDT \triangleq \limpf \frac{T^\star}{\nicefrac{1}{\log P}}  = \limpf T^\star \log P.
				}			
			Let us introduce also the Degrees-of-Freedom\cite{Davoodi2016_TIT_DoF} (DoF) metric, which are defined as $\DoF \triangleq \limpf \frac{C(P)}{\log P}$, where $C(P)$ denotes the capacity of the setting. 
			Hence,  the $\NDT$ can also be written as $$\NDT = \frac{K(1-\gamma)}{\DoF}.$$

			The metrics presented above are similarly  defined  when considering only uncoded prefetching, in which case we will just insert a sub-index $u$ (e.g. $\NDT_u,T_{u,\varepsilon,\phi}$). 
				%
			Furthermore, when necessary, we will use the full notation 
			$\NDT(K,L,\gamma,K_P)$ to reflect the particular network configuration. 
			For example, the result for the well-known $K$-user SISO BC with uncoded prefetching would correspond to $\NDT_u(K,1,\gamma,0)$. 

		\section{Main results}\label{se:main}
		We will henceforth make use of the term $(K,L,\gamma,K_P)$ MISO BC to denote the $L\times K$ cache-aided MISO BC in which only a fixed set of $K_P\leq L$ users provide perfect CSIT, whereas the other $K_F=K-K_P$ users provide finite precision CSIT. 
		We begin with the characterization of the optimal rate-memory trade-off for the particular case of uncoded placement. 
			\begin{theorem}\label{theo:claim1}
				Under the assumption of  uncoded prefetching, 
				the optimal normalized delivery time of the $(K,L,\gamma,K_P)$~MISO BC when $K_P\leq L$ is given by\footnote{For sake of readability and concision, we omit  the fact that the $\NDT$ is naturally upper bounded by $(1-\gamma)$; for example, in~\eqref{eq:theo1_1}, it holds that $\NDT_u(K,L,\gamma, K_P) = \min \big((1-\gamma),\Conv_{(K_F+1)\gamma}\big(\frac{(K_F+1)(1-\gamma)}{(K_F+1)\gamma + 1}\big)\big).$}
				\eqm{
					\NDT_u(K,L,\gamma, K_P) 
						  =  \Conv_{(K_F+1)\gamma}\Big(\frac{(K_F+1)(1-\gamma)}{(K_F+1)\gamma + 1}\Big),\label{eq:theo1_1}
				}
				where $K_F = K - K_P$ and $\Conv_A\big(f(A)\big)$ denotes the lower convex envelope of the points $\big\{\big(A,\ f(A)\big) \,$ $|\, A \in \{0, 1, . . . , K\}\big\}$. This implies that 
				\eqm{
					\NDT_u(K,L,\gamma, K_P) 
						& = \NDT_u(K_F+1,1,\gamma,0).\label{eq:theo1_1bb}
				}
			\end{theorem}	
				\begin{proof}
					The achievable scheme and the converse are presented in Section~\ref{se:achiev} and Section~\ref{se:proof_theorem1_single}, 
					respectively. 
				\end{proof}
			
			\begin{remark}\label{lem:coro1b}
					Theorem~\ref{theo:claim1} shows that the optimal normalized delivery time required to serve $K_F+K_P$ ($K_P\leq L$) users in the multi-antenna TX setting is the same as the one required to serve $K_F+1$ users in the standard single-antenna TX setting. 
					In other words, starting from a cache-aided setting with $K_F$ users and only finite precision CSIT, we can add as many as $L$ users who provide perfect CSI at the cost of a single finite-precision-CSIT user. 
			\end{remark}			
						
			\begin{remark}\label{lem:coro1baa}
				The $\NDT$ in~\eqref{eq:theo1_1} is not achievable through separate transmission to the two classes of users, since that approach attains a $\NDT$ of $\frac{K_F(1-\gamma)}{1+K_F\gamma} + (1-\gamma)$, which is strictly bigger. 
			\end{remark}			

			\begin{remark}\label{lem:coro1bss}
					Theorem~\ref{theo:claim1} also implies that perfect CSIT for a single user does not improve the $\NDT$ with uncoded prefetching with respect to the case where all  users provide only finite precision CSIT. This is analogous to the collapse of DoF proved in~\cite{Davoodi2016_TIT_DoF} for the cacheless ``PN'' setting with perfect CSIT for only one user.   			
			\end{remark}			
		
		Furthermore, we can also write the expression in~\eqref{eq:theo1_1} as
			\eqm{
				\frac{(K_F+1)(1-\gamma)}{(K_F+1)\gamma + 1}
						 =  \frac{K(1-\gamma)}{K\gamma + \frac{K}{K-K_P+1}},
				}		
		which allows us to compare our setting to the standard cache-aided MISO BC with perfect CSIT for all the users, whose equivalent expression is  $\frac{K(1-\gamma)}{K\gamma + L}$. 
		
		Next, we remove the assumption of uncoded prefetching, and we show that the proposed scheme  is within a factor of 2.01 from the optimal.		
			\begin{theorem}\label{theo:claim1b}
				The optimal normalized delivery time of the $(K,L,\gamma,K_P)$~MISO BC when $K_P\leq L$ satisfies that   
					\eqm{
						&\NDT_u(K_F+1,\,1,\,\gamma,\,0) \geq
						  \NDT(K,L,\gamma,K_P) 
						\geq \frac{1}{2.00884}\NDT_u(K_F+1,\,1,\,\gamma,\,0) .
					}
			\end{theorem}	
				\begin{proof}
					The achievable scheme is the same as for Theorem~\ref{theo:claim1},  
					whereas the converse is presented in Section~\ref{se:proof_theo1b}. 
				\end{proof}		
		
			While the achievable scheme can be directly generalized for the case where $K_P > L$, the generalization of the bounds is not straightforward. Nevertheless, it is expected that the insight in Remark~\ref{lem:coro1b} will hold, i.e., that we can add $K_P$ perfect-CSIT users at the cost of $\lceil K_P/L\rceil$ finite-precision-CSIT users.

		\section{Achievability of Theorem~\ref{theo:claim1} and Theorem~\ref{theo:claim1b}}\label{se:achiev}
		We present the scheme achieving the $\NDT$ of Theorem~\ref{theo:claim1} and Theorem~\ref{theo:claim1b} for the case where $(K_F+1)\gamma\in\{0,1,\dots,K_F+1\}$, since for the general case the $\NDT$ can be achieved through the standard memory-sharing techniques. 
		
		We denote the set of users providing perfect CSI as $\Kc_P \subseteq [K]$. 
		Our aim is to provide the users belonging to $\Kc_P$ with both full spatial multiplexing gains and full coded caching gains, which will naturally surpass the performance of a simple separate transmission to the two sets of users. 
		As it turns out, the way to achieve these full gains is allocating identical cache content to all users in $\Kc_P$, and thus the scheme assumes only $\Lambda \triangleq K - K_P + 1$ different cache states. This is clarified below. 		

		
		%
			
			\subsection{Placement}\label{se:plac}
			We consider  $\Lambda$ cache states of size $MB$ bits, and denote them as $Z^{(c)}_i$, for $i\in[\Lambda]$. 
			Let $\Lambda\gamma$ be an integer, and $\Tc$ be the set of $\binom{\Lambda}{\Lambda\gamma}$ subsets in $[\Lambda]$ of size $\Lambda\gamma$, i.e.,  
						$\Tc \triangleq \{\tau\subset [\Lambda] : |\tau|=\Lambda\gamma\}.$
			First, we split each message $W_n$, $n\in[N]$, into $|\Tc|\triangleq \binom{\Lambda}{\Lambda\gamma}$ non-overlapping subfiles of equal size, such that
						$W_n \to \{W_{n,\tau} : \tau \in \Tc\}$,
				and we assign the content of each of the $\Lambda$ caches as
							$Z^{(c)}_i = \{W_{n,\tau} : i\in\tau,\ \tau\in\Tc\}_{n=1}^N$. 
			Although the subpacketization is analogous to the one developed by Maddah-Ali and Niesen in \cite{MaddahAli2014}, in our case the cache assignment is different: 
			Let us first assume that the users providing perfect CSI are the last ones, i.e., $\Kc_P =  \{K-K_P+1,\dotsc,K\}$. Then, all the users in $\Kc_P$ store the content of the last cache state,  $Z_{i}=Z^{(c)}_{\Lambda}$, $\forall i\in\Kc_P$, 
			while each of the remaining users stores a different cache state, such that $Z_i = Z^{(c)}_i$, $\forall i\in[\Lambda-1]$.
			Hereinafter, we will refer to the set of users that share the same cache as a \emph{cache group}, even if only one of these groups has more than one user.  
		
							
			\subsection{Transmit Signal}\label{se:trans}
			During the delivery phase, the TX simultaneously serves  $\Lambda\gamma + 1$ cache groups. Let 
					$\Xc \triangleq \{\chi\subseteq[\Lambda] : |\chi| = \Lambda\gamma + 1\}$
			be defined as the set of $|\Xc|= \binom{\Lambda}{\Lambda\gamma+1}$ subsets of size $\Lambda\gamma+1$ in $[\Lambda]$. 
			The delivery consists in $|\Xc|$ sequential transmissions, one for each $\chi \in\Xc$. 
			Let $W_{d_j,\chi\backslash c_j}$ denote the part of the file requested by user $j$ that is stored in all the caches in $\chi$ except in the cache allocated at user~$j$ ($c_j$). Note that $c_j = j$ if $j\notin \Kc_P$, and $c_j = \Lambda$ if $j\in \Kc_P$. 
			For each possible~$\chi\in\Xc$, the transmit signal takes a different expression depending of whether $\Lambda\in \chi$ or not (recall that all users in $\Kc_P$ store the $\Lambda$-th cache state, which is the only one shared by several users). 

				\subsubsection{Case $\Lambda\notin \chi$}\label{se:transB}
				The transmit signal  when the $\Lambda$-th cache group is not in $\chi$ is given by 
					\eqm{
							\bX(t) = \vv_{0}(t)\sum\nolimits_{\smash{g\in\chi}} W_{d_g,\chi\backslash g},
					}
				where $\vv_{0}(t)$ is a randomly chosen $L\times 1$  precoding vector.
				\subsubsection{Case $\Lambda\in \chi$}\label{se:transA}
				In this case, we have that
					\eqm{
							\bX(t) =  \vv_{0}(t)\sum_{\mathclap{g\in\{\chi\backslash \Lambda\}}} W_{d_g,\chi\backslash g} + \sum_{p\in \Kc_P} \vv_{\Kc_P\backslash p}(t)W_{d_{p}, \chi\backslash \Lambda},
					}			
				where $\vv_{\Kc_P\backslash p}(t)$ is an $L\times 1$  precoding vector designed to belong to the null space of the $K_P-1$ users in $\{\Kc_P\backslash p\}$. 
				
			\subsection{Received Signal}\label{se:Rec}
			
				\subsubsection{Received signal if $\Lambda\notin \chi$}\label{se:recB}
				In this case, the received signal at user~$i$ is given by 
					\eqm{\label{}
							Y_{i}(t) = \hv_i(t)\vv_{0}(t)\Big(\sum\nolimits_{g\in\chi} W_{d_g,\chi\backslash g}\Big) + \zeta_{i}(t).
					}
				Note that, for any $i\in\chi$, user~$i$ can remove all the undesired messages because they are cached in $Z_i$. 
				Thus, user~$i$ obtains
					\eqm{
							Y'_{i}(t) = \hv_i(t)\vv_{0}(t) W_{d_i,\chi\backslash i} + \zeta_{i}(t),
					}				
				and it can decode the intended subfile $W_{d_i,\chi\backslash i}$. 
				
				\subsubsection{Received signal if $\Lambda\in \chi$}\label{se:recA}
				Any user not in $\Kc_P$ can decode its subfile as in the previous case, since all the non-intended subfiles are available in its local cache. 
				For users in $\Kc_P$, 
				they can  remove the messages intended by the finite-precision-CSI users, which are available in their local cache, thus obtaining
					\eqm{
							Y'_{i}(t) = \hv_i(t)\Big(\sum\nolimits_{p\in \Kc_P} \vv_{\Kc_P\backslash p}(t)W_{d_{p}, \chi\backslash \Lambda}\Big) + \zeta_{i}(t).
					}				
				By definition of $\vv_{\Kc_P\backslash p}(t)$, it holds that $\hv_i(t)\vv_{\Kc_P\backslash p}(t) = 0$ for any $i\in\Kc_P : i\neq p$. 
				From that, it follows that 
					\eqm{
							Y'_{i}(t)  = \hv_i(t)\vv_{\Kc_P\backslash i}(t) W_{d_{i}, \chi\backslash \Lambda} + \zeta_{i}(t),
					}		
				and thus, user~$i$ can decode its intended subfile $W_{d_i,\chi\backslash \Lambda}$.

			\begin{remark}\label{lem:coro1bssaa}
					The users must decide whether they will provide perfect CSIT already at the placement phase to be able to correctly assign the cache states. 
					Yet, it is not difficult to see that the optimal $\NDT$ is also achieved if only a subset $\Kc_P'\subset\Kc_P$ of users in $\Kc_P$ appear for delivery. 
			\end{remark}			
			
			For the case where $K_P > L$, the scheme is easily generalized by having $\lceil K_P/L\rceil$ cache groups with several users.

			\subsection{Degrees-of-Freedom}\label{se:dof}
			This scheme  always serves $\Lambda\gamma + 1$ groups simultaneously.
			However, the number of served users varies depending on whether $\Lambda\in \chi$ or not. 
			Concretely, the TX simultaneously serves  
				\eqm{\label{eq:dof-cases}
					\begin{cases}
							\Lambda\gamma + 1 \text{ users} & \text{if $\Lambda\notin \chi$}\\
							\Lambda\gamma + K_P \text{ users} & \text{if $\Lambda\in \chi$}. 
					\end{cases}
				}
			Let $\Xc_{\Lambda}\triangleq \{\chi\in \Xc :\Lambda \in \chi\}$ denote the set of those $\chi\in\Xc$ that include $\Lambda$. 
			Then, it follows that 
				\eqm{\label{eq:chi-size}
					|\Xc_{\Lambda}| = |\{\chi\in\Xc : \Lambda\in\chi\}| = \binom{\Lambda-1}{\Lambda\gamma}.
				}
			Since all the subfiles have the same size, the time required for decoding each of them is the same, no matter which user-set $\chi$~is being served. 
			Thus, from~\eqref{eq:dof-cases} and~\eqref{eq:chi-size} it follows that
				\eqm{
					\DoF 	&= \frac{\binom{\Lambda-1}{\Lambda\gamma}\LB \Lambda\gamma + K_P\RB + \Big(\binom{\Lambda}{\Lambda\gamma+1}-\binom{\Lambda-1}{\Lambda\gamma}\Big)\LB \Lambda\gamma + 1\RB}{\binom{\Lambda}{\Lambda\gamma+1}} \notag \\[1ex]
								&= \LB 1 + \Lambda\gamma\RB + \LB K_P-1\RB\frac{\Lambda\gamma+1}{\Lambda}. \label{eq:dof_1}
				}
			Therefore, by applying $\Lambda= K-K_P+1$ into~\eqref{eq:dof_1}, we obtain 
				\eqm{\label{eq:dof_eq_1}
					\DoF &= 1 + K\gamma + \frac{K_P-1}{K-K_P+1}.
				}			
			Since the $\NDT$ can be written as $\NDT = \frac{K(1-\gamma)}{\DoF}$, by applying that $\Lambda= K-K_P+1=K_F+1$ into~\eqref{eq:dof_eq_1} yields 
				\eqm{
					\NDT 
						 = \frac{\Lambda(1-\gamma)}{1+\Lambda\gamma} = \NDT_u(K_F + 1,1,\gamma, 0), \label{eq:last_eq_achiev}
				}			
			which concludes the proof of the achievability of both theorems for the cases in which $\Lambda\gamma = (K_F+1)\gamma$ is an integer. In other cases, the NDT can be achieved through the usual memory-sharing approach\cite{MaddahAli2014,Yu2018_TIT,Yu2019_TIT}. 
		 
		\section{Converse of Theorem~\ref{theo:claim1} and Theorem~\ref{theo:claim1b}}\label{se:proof_theo1}
		We first present two useful lemmas that are instrumental  for the converse of both theorems. 
		This will be followed first by the converse of Theorem~\ref{theo:claim1} and later 
		by that of Theorem~\ref{theo:claim1b}.\vspace{.35ex} 
		
			\subsection{Lower Bounding the Achievable Delivery Time}\label{se:lemma_entropy}\vspace{.5ex}
			In the following, we derive a lower bound for any {$\varepsilon$-achievable} delivery time.  
						Before presenting the result, we introduce the notations $\Zc_{[j]}\triangleq \{Z_i\}_{i\in[j]}$,  $\Wc_{[j]}\triangleq \{W_{d_i}\}_{i\in[j]}$, and  $\Hc\triangleq \{h\sij(t)\}_{i\in[K],j\in[L],t\in[\tau]}$. 		
We start by focusing on the ``PN'' problem with side information. 
			This important setting can be characterized by the following key lemma. 
			
			\begin{lemma}\label{lem:ais1}
				Consider the $(K,L,\gamma,K_P=1)$~MISO BC where the TX has perfect CSI only for one user. 
				Then,  it holds that
					\eqm{\label{eq:kem_ais_1}
						&H(Y_k\Tt| \Zk,\Wk,\Hc) - H(Y_{k+1}\Tt| \Zk,\Wk, \Hc) 
							\geq \tau\, o(\log P),\qquad \forall k\in[K].
					}
			\end{lemma}	
				\begin{proof}
					The proof follows from a new application of the Aligned Image Set approach, and it is presented in Appendix~\ref{se:proof_ais1}. 
				\end{proof}		
			The previous lemma is instrumental in the derivation of our next lower bound on the delivery time. 
			
			\begin{lemma}\label{lem:lower1}
				Consider the $(K,L,\gamma,K_P=1)$~MISO BC where the TX has perfect CSI only for one user. 
				For any prefetching scheme $\phi$ and any demand $\bd$, the $\varepsilon$-achievable delivery time $T_{\varepsilon,\phi}$ is lower-bounded by
					\eqm{\label{eq:kem_low_1}
						T_{\varepsilon,\phi}\big(\log (1+KP) + o(\log P)\big)
						& \geq \frac{1}{B}\sum_{k=1}^{\min(N,K)}  H (W_{d_k} \mid \Zc_{\ang{k}}, \Wc_{[k-1]}) -\min(N,K) \Big(\frac{1}{B} + \varepsilon\Big).\vspace{2ex}
					}
			\end{lemma}	
			\begin{proof}
								
				Let us consider that a particular delay $T_{\varepsilon,\phi}$ is $\varepsilon$-achievable. 
				Then, for any request $\bd = \{{d_1},{d_2},\dotsc,{d_K}\}$, there exists a transmitted signal vector $\bX_d\Tt$ 
				such that each user~$k\in[K]$ can decode $W_{d_k}$ from $Z_k$ and $Y_k\Tt$ with probability of error at most $\varepsilon$.  
				From Fano's inequality and the fact that conditioning reduces entropy, it follows that
					\eqm{
						H(W\sdk\mid Y_k\Tt, \Zk,\Wkm, \Hc) \leq  1 + \varepsilon B,\  
					}
				for any $k\in[K]$. From the above, it follows that 
					\eqm{
						&H(Y_k\Tt\mid \Zk,\Wkm,\Hc) \geq  H(W\sdk\mid \Zk,\Wkm,\Hc) 
						+ H(Y_k\Tt\mid \Zk,\Wk,\Hc) - (1 + \varepsilon B), 
					}	
				for any $k\in[K]$. 
				If we sum up the previous inequality for all $k\in[\minkn]$, where we have introduced the notation $K'\triangleq\min(K,N)$, we have that 
					\eqm{
						\sum_{k=1}^{\minkn} H(Y_k\Tt\mid \Zk,\Wkm, \Hc) &\geq  \sum_{k=1}^{\minkn} \Big(\  H(W\sdk\mid \Zk,\Wkm, \Hc) \Big. \notag \\
						& \quad\qquad\ \ \Big. {}+{} H(Y_k\Tt\mid \Zk,\Wk, \Hc)  -(1 + \varepsilon B)\Big). \label{eq:sum_ineq1}
					}		
				By re-ordering terms in~\eqref{eq:sum_ineq1}, we obtain 	
					\eqm{
						H(Y_1\Tt\mid Z_1, \Hc) 
								& \geq  \sum_{k=1}^{\minkn} H(W\sdk\mid \Zk,\Wkm, \Hc) - \minkn(1 + \varepsilon B) \nonumber \\
								& \ \quad +  \sum_{k=1}^{\minkn-1} H(Y_k\Tt\mid \Zk,\Wk, \Hc) - H(Y_{k+1}\Tt\mid \Zc_{[k+1]},\Wk, \Hc)\\
								& \myoverset{(a)}{\geq}  \sum_{k=1}^{\minkn} H(W\sdk\mid \Zk,\Wkm, \Hc) - \minkn(1 + \varepsilon B) \nonumber \\
								& \  \quad +  \sum_{k=1}^{{\minkn-1}} H(Y_k\Tt\mid \Zk,\Wk, \Hc) - H(Y_{k+1}\Tt\mid \Zk,\Wk, \Hc),  \label{eq:ais_prev_q} 
					}		
				where $(a)$ comes from the fact that conditioning reduces entropy. 
				Now, we upper bound the LHS of the above inequality as $H(Y_1\Tt\mid Z_1, \Hc)\leq \tau\log (1+KP)$, and we lower bound the RHS of~\eqref{eq:ais_prev_q} by applying Lemma~\ref{lem:ais1}, which implies that
					\eqm{
						 \tau\log (1+KP) 
								& \geq  \sum_{k=1}^{\minkn} H(W\sdk\mid \Zk,\Wkm, \Hc) - \minkn(1 + \varepsilon B) - \tau\, o(\log P). \label{eq:proof_lem_last1}
					}
				We can remove the conditioning on $\Hc$ because both messages and caches are independent of the channel.  
				Since $\tau =T_{\varepsilon,\phi}\,B$, we obtain Lemma~\ref{lem:lower1} by rearranging the terms in~\eqref{eq:proof_lem_last1}. 
			\end{proof}	
			
			Lemma~\ref{lem:ais1} and Lemma~\ref{lem:lower1} are the key contribution of this work, and they are essential for proving both Theorem~\ref{theo:claim1} and Theorem~\ref{theo:claim1b}. Lemma~\ref{lem:lower1} can be seen from two perspectives: First, it represents the non-trivial extension of \cite[Lemma 2]{Yu2019_TIT} from the single-server, error-free, shared link scenario to the $(K,L,\gamma,K_P)$~MISO BC 
			with $K_P=1$. 
			%
			As explained in \cite[Section V]{Yu2019_TIT}, these lemmas represent an enhanced cut-set bound that improves the compound cut-set bound previously used in other works\cite{MaddahAli2014}. 
			Second, it extends the results of the ``PN'' BC setting to the important case where there exists side information at the receiver side. 
		
			\subsection{Proof of Theorem~\ref{theo:claim1}}\label{se:proof_theorem1_single}
			We now proceed to derive the converse for the optimal $\NDT$ under uncoded prefetching presented in Theorem~\ref{theo:claim1}.  
			As a first step, let us restrict ourselves to the case $K_P=1$, and let $K'\triangleq\min(N,K)$. 
			Before applying the derivation, recall that the bits in the library are i.i.d. and uniformly distributed. 
			Let $B_{n,b}$ denote the $b$-th bit of file $n$, and let $\Bc^{(k)}_{d_k,b}$ represent the event that $B_{d_k,b}$ is not cached by any user in the set~$\{i\}_{i=1}^k$. 
			Then, it follows (cf.\cite{Yu2018_TIT}) that 
			$$\sum_{k=1}^{\minkn} H (W_{d_k} \mid \Zc_{\ang{k}}, \Wc_{[k-1]}) \geq \sum_{k=1}^{\minkn}  \sum_{b=1}^{B} \Ind\big( \Bc^{(k)}_{d_k,b}\big),$$   
			where $\Ind(\cdot)$ denotes the indicator function. 
			Thus, from Lemma~\ref{lem:lower1}, and by applying the same steps as in~\cite[Appendix~A]{Yu2018_TIT}, the value of $T^\star_{u,\varepsilon,\phi}$ is lower bounded by 
				\eqm{\label{eq:kem_low_2}
					 T^\star_{u,\varepsilon,\phi}\big(\log (1+KP) + o(\log P)\big) 
							 \geq  \Conv_{K\gamma}\Big( \frac{K(1-\gamma)}{K\gamma + 1}\Big)  
							 -\minkn \Big(\frac{1}{B} + \varepsilon\Big). \notag 
				}
			Since 
					$T_{u}^\star = \sup_{\varepsilon>0} \limsup_{B\rightarrow\infty}\min_{\phi} T^\star_{u,\epsilon,\phi}$, 
			it follows that 	
				\eqm{
					T^\star_u  \big(\log (1+KP) + o(\log P)\big)
						&  = \Conv_{K\gamma}\Big( \frac{K(1-\gamma)}{K\gamma + 1}\Big). 
				}
			Now, applying  the fact that $\NDT_u$ is defined as $\NDT_u = \limpf T^\star_u\log P$ yields 
				\eqm{
					\NDT_u(K,L,\gamma, 1)  =\Conv_{K\gamma}\Big( \frac{K(1-\gamma)}{K\gamma + 1}\Big), \label{eq:proof_P_1}
				}		
			and thus yields that $\NDT_u(K,L,\gamma, 1) = \NDT_u(K,1,\gamma,0)$, which proves Theorem~\ref{theo:claim1} for the case $K_P=1$.  
			
			Let us now consider the general case in which $K_P\leq L$. 
			We start by considering the reduced setting in which we serve  the $K - K_P$ users for which there is finite precision or no CSIT and only one additional user among the $K_P$ users providing perfect CSIT. 
			The optimal delay on this reduced scenario will clearly lower bound the optimal delay of the whole setting. 
			This 
			optimal delay of the reduced scenario is given by~\eqref{eq:proof_P_1} after setting the total number of users to $ K - K_P+1$. Then, it follows from~\eqref{eq:proof_P_1} that 
				\eqm{
					\NDT_u(K,L,\gamma, K_P) 
							& \geq \NDT_u(K - K_P + 1,L,\gamma, 1) \\
							& = \NDT_u(K - K_P + 1,1,\gamma,0),
				}
			which concludes the proof of Theorem~\ref{theo:claim1}. \qed

			\subsection{Proof of Theorem~\ref{theo:claim1b}}\label{se:proof_theo1b}		
			From the derivation of Lemma~\ref{lem:lower1}, the converse of Theorem~\ref{theo:claim1b} follows a similar path as the proof of~\cite[Theorem~2]{Yu2019_TIT} for the error-free symmetric setting (see~\cite[Section V]{Yu2019_TIT}). 
			The proof is composed of two main steps. 
			First, we obtain a symmetric version of the lower bound in Lemma~\ref{lem:lower1}. 
			After that, we minimize the symmetric bound for any prefetching scheme.

			We first consider the case in which $K_P = 1$, as in Lemma~\ref{lem:lower1}. 
			We start by applying a symmetrization over all the possible requests and user orderings on the RHS of~\eqref{eq:kem_low_1}. 			
			For that, we define $\eta : [N] \rightarrow [N]$ as a permutation over  the $N$ files in the library, and $\Sc_N$ as the set of all such $N!$ permutations. 
			In a similar way, we define $\kappa : [K] \rightarrow [K]$ as a permutation over the set of users and $\Sc_K$ as the set of all such $K!$ permutations.
			Given a specific pair of permutations $\eta$ and $\kappa$ in $\Sc_N\times\Sc_K$, a certain user $\kappa(k)$ requests a file $W_{\kappa(k)} = \eta(d_{\kappa(k)})$. 
			We define $\Zc_{\kappa([a])} \triangleq \{Z_{\kappa(i)}\}_{i\in[a]}$ and $\Wc_{\kappa([a])} \triangleq \{W_{\kappa(i)}\}_{i\in[a]}$. 
			Now, we consider the average of~\eqref{eq:kem_low_1} from Lemma~\ref{lem:lower1}	over all possible permutations in $\Sc_N\times\Sc_K$, which leads to
				\eqm{\label{eq:kfem_low_1}
					T_{\varepsilon,\phi}\big(\log (1+KP) + o(\log P)\big) 
						& \geq \frac{1}{K!N!B}\sum_{(\eta,\kappa)\in\Sc_N\times\Sc_K}\sum_{k=1}^{\minkn} H (W_{\kappa(k)} \mid \Zc_{\kappa([k])}, \Wc_{\kappa([k-1])}) \notag\\
						& \quad -\minkn (\frac{1}{B} + \varepsilon).
				}		
			Let us now transcribe Lemma~3 in~\cite{Yu2019_TIT}, which we will apply in the following. 
				\begin{lemma}[{\cite[Lemma 3]{Yu2019_TIT}}]\label{lem:yu_lemma}
						For any parameters $s\in[\minkn]$, $\alpha\in[0,1]$, and prefetching scheme $\phi$, we have that
					\eqm{
						&\frac{1}{K!N!B}\sum_{(\eta,\kappa)\in\Sc_N\times\Sc_K}\sum_{k=1}^{\minkn} H (W_{\kappa(k)} \mid \Zc_{\kappa([k])}, \Wc_{\kappa([k-1])}) \notag\\
						&\hspace{37ex}\geq 	s-1-\alpha - \frac{s(s-1) - \ell(\ell-1) + 2\alpha s}{2(N-\ell-1)}M	\label{eq:proof_1a00}, 
					}
				where $\ell\in[s]$ is the minimum value satisfying 
					\eqm{
							\frac{s(s-1) - \ell(\ell-1)}{2}  + \alpha s \leq (N-\ell+1)\ell. 
					}
				\end{lemma}
				
			We recall that the optimal delivery time has been defined in~\eqref{eq:def_tstar} as $T_{}^\star \triangleq \sup_{\varepsilon>0} \limsup_{B\rightarrow\infty}\min_{\phi} T_{\epsilon,\phi}$. 
			Hence, we can apply~Lemma~\ref{lem:yu_lemma} into~\eqref{eq:kfem_low_1} 
			to obtain that		
				\eqm{
					& T^\star\big(\log (1+KP) + o(\log P)\big)  \geq 	s-1-\alpha  - \frac{s(s-1) - \ell(\ell-1) + 2\alpha s}{2(N-\ell-1)}M	\label{eq:proof_1a}.
				}
			We can take the limit of~\eqref{eq:proof_1a} as $P$ approaches infinity to obtain the $\NDT$, which leads to 
				\eqm{
					\NDT \geq 	s-1-\alpha - \frac{s(s-1) - \ell(\ell-1) + 2\alpha s}{2(N-\ell-1)}M	\label{eq:proof_1bbb}.
				}						
			Let us make explicit the dependence of the $\NDT$ on the parameters $(K=K_F+1,L,\gamma,K_P=1)$. 				
			We have shown in Section~\ref{se:achiev} that we can achieve a $\NDT$ equal to	$\NDT_u(K_F+1,1,\gamma,0)$ (see~\eqref{eq:last_eq_achiev}).  
			From~\cite{Yu2018_TIT}, we also know that $\NDT_u(K_F+1,1,\gamma,0)$ is upper-bounded by the exact minimum delivery time required for decentralized caching with uncoded prefetching, which we denote by $\NDT_{u,\dec}(K_F+1,1,\gamma,0)$. Then, we can write (cf.~\cite{Yu2018_TIT}) that
				\eqm{
						\NDT_u(K_F+1,1,\gamma,0) \leq \NDT_{u,\dec}(K_F+1,1,\gamma,0) \triangleq \frac{N-M}{M}\LB 1-(1-M/N)^K\RB. 
				}
			From \cite[Lemma 1]{Yu2019_TIT} and \cite[Appendix G]{Yu2019_TIT}, it follows that
				\eqm{
					 \NDT(K_F+1,L,\gamma,1) & \geq 	s-1-a - \frac{s(s-1) - \ell(\ell-1) + 2as}{2(N-\ell-1)}M	\label{eq:proof_1e}\\
						& \geq 	\frac{1}{2.00884}\NDT_{u,\dec}(K_F+1,1,\gamma,0) \label{eq:proof_1f}\\
						& \geq 	\frac{1}{2.00884}\NDT_u(K_F+1,1,\gamma,0) \label{eq:proof_1fb}\\
						& = \frac{1}{2.00884}\Conv_{(K_F+1)\gamma}\LB \frac{\binom{K_F+1}{(K_F+1)\gamma+1}}{\binom{K_F+1}{(K_F+1)\gamma}}\RB \\
						&	=  \frac{1}{2.00884}\Conv_{(K_F+1)\gamma}\LB\frac{(K_F+1)(1-\gamma)}{(K_F+1)\gamma+1}\RB \\
						& = \frac{1}{2.00884}\NDT_u(K_F+1,\,1,\,\gamma,\,0).   \label{eq:proof_1g}
				}
			Hence, 
				\eqm{
					\frac{1}{2.00884}\NDT_u(K_F+1,\,1,\,\gamma,\,0) \leq \NDT(K_F+1,L,\gamma,1)  \leq \NDT_u(K_F+1,\,1,\,\gamma,\,0).
				}	
			Note that, for the regime where $K_P\leq L$, we know from Section~\ref{se:achiev} that we can achieve a $\NDT$ equal to	$\NDT_u(K_F+1,1,\gamma,0)$. 
			On the other side, it holds that $\NDT(K,L,\gamma,K_P) \geq \NDT(K_F+1,L,\gamma,1)$, since we can obtain the RHS of the inequality by considering that $K_P-1$ users in $\Kc_P$ do not require any information. Then, it follows from~\eqref{eq:proof_1g} that
				\eqm{
					\frac{1}{2.00884}\NDT_u(K_F+1,\,1,\,\gamma,\,0) \leq \NDT(K,L,\gamma,K_P)  \leq \NDT_u(K_F+1,\,1,\,\gamma,\,0),
				}	
			which concludes the proof of Theorem~\ref{theo:claim1b}. \qed

		\section{Conclusions}\label{se:conclusions}
		We have considered the cache-aided $L\times K$ MISO BC where only a fixed set of $K_P\leq L$ users provide perfect CSI to the transmitter, whereas the remaining users provide only finite precision CSIT. 
		This setting also corresponds to the side-information variant of the well-known ``PN'' BC setting. 
		For this setting, we have derived the optimal rate-memory trade-off at high SNR under uncoded placement, and characterized the same trade-off within a factor of~2.01 for coded placement. 
		The proposed scheme capitalizes on the fact that the optimal trade-off cannot be achieved through separate transmission between the users that provide different CSIT level. 
		The derived limits clearly show that coded-caching and multi-antenna gains are synergistic and schemes should integrate both aspects, but also that the existence of users that provide only finite precision CSIT greatly reduces the performance, which is in line with the results of the ``PN'' setting without side information.
		Analyzing the case $K_P > L$ is a meaningful extension currently under investigation, and finally, considering partial or heterogeneous CSI settings are also interesting research directions. \vspace{4ex}

													
\begin{appendices}%
	\renewcommand{\thesectiondis}[2]{\Alph{section}:}
		
		\section{Proof of Lemma~\ref{lem:ais1}}\label{se:proof_ais1}
		We  make use of the Aligned Image Set (AIS) technique introduced by Davoodi and Jafar in \cite{Davoodi2016_TIT_DoF}, and, more specifically, the proof builds on the results in \cite{Davoodi2020_sumset}. 
		We demonstrate Lemma~\ref{lem:ais1} for real channels, and the extension to complex channels, while cumbersome, follows the same steps as in~\cite{Davoodi2016_TIT_DoF,Davoodi2018_TIT_BC,Davoodi2020_sumset}. 
		First let us define some notation that is generally used in the AIS approach.\footnote{We have slightly changed the notation in this paper in order to avoid ambiguous notation with respect to the rest of the document. %
										In the AIS literature, the set of random variables $h\in\Hc$ is normally denoted $g\in\Gc$, and the set of finite constants $a\sji$ as is normally denoted  $h\sji$. Moreover, the linear combinations $F_j$, $F^b_j$, are normally denoted as $L_j$, $L^b_j$, which would be ambiguous with the number of transmit antennas $L$.}

			\begin{definition}
				For real numbers $x_1,x_2,\dots,x_L$, define the notations $F^b_j(\{x_i\mid i\in[L]\})$, $F_j(\{x_i \mid i\in[L]\})$ as 
					\eqm{
							F^b_j(x_1,x_2,\dots,x_L) &\triangleq \sum_{i\in[L]}\lfloor h_{j,i}x_i\rfloor, \\
							F_j(x_1,x_2,\dots,x_L) &\triangleq \sum_{i\in[L]}\lfloor a_{j,i}x_i\rfloor,
					}
				for distinct random variables $h_{j,i}\in\Hc$ satisfying the bounded density assumption in Definition~\ref{def:bounded}, and for some arbitrary real valued and finite constants $a_{j,i}\in\Hc$, $|a_{j,i}|\leq \delta_z < \infty$. 
				The subscript $j$ is used to distinguish among multiple sums. 
			\end{definition}		
		In order to apply the AIS approach, the first step consists on discretizing the channel to obtain a deterministic channel model. 
		The deterministic channel model is defined by the outputs
			\eqm{
				\Ybar_k(t) = \sum_{i=1}^L \lfloor h_{k,j}(t) \Xbar_j(t)\rfloor,\qquad \forall k\in[K],
			}
		where the discretized inputs are defined such that $\Xbar_j(t)\in\{0,1,\dots,\Pb\}$, $\forall t\in\Nb$, and where $\Pb\triangleq \lfloor\sqrt{P}\rfloor$. 
		Then, it is known that this deterministic channel model outer bounds the initial model~\cite{Davoodi2016_TIT_DoF,Davoodi2018_TIT_BC,Davoodi2020_sumset}. 	

		Since Lemma~\ref{lem:ais1} is restricted to the case with only one user providing perfect CSIT, let us consider w.l.o.g. that user~$K$ is the user for which there is perfect CSIT. 
		Then, it holds that
			\eqm{
				\Ybar_k(t) & = F^b_k(t)\LB \Xbar_1(t), \dots, \Xbar_L(t)\RB, \quad  \forall k\in[K]\backslash K, \label{eq:y_is_la}\\
				\Ybar_K(t) & = F(t)\LB \Xbar_1(t), \dots, \Xbar_L(t)\RB. \label{eq:y_is_lb}
			}		
		
		Lemma~\ref{lem:ais1} follows from \cite[Theorem~3]{Davoodi2020_sumset}. 
		We present below a simplified version of~\cite[Theorem~3]{Davoodi2020_sumset} tailored for our particular case. 
			\begin{theorem}[{\cite[Theorem~3]{Davoodi2020_sumset}}]\label{theo:theorem_ais}
				Consider random variables $\Xbar_j(t)\in\Xc$, $j\in [L]$, $t\in\Nb$, where $\Xc \triangleq \{0,1,2,\dots,\Pb\}$.
				Let those random variables $\Xbar_j(t)$ be independent of $\Hc$. 
				Define 
					\eqm{
						A(t) & = F^b(t)\big(\Xbar_1(t),\dots,\Xbar_L(t)\big), \\
						B(t) & = F(t)\big(\Xbar_1(t),\dots,\Xbar_L(t)\big).
					}
				Let $\Hc(A)\subset\Hc$ denote the set of all bounded density channel coefficients that appear in $A\Tt  = \{F^b(t)\big(\Xbar_1,\dots,\Xbar_L\big)\}_{t\in[\tau]}$.  
				Let $S$ be a random variable such that, conditioned on any $\Hc_0 \subset \{S\}\cup \{\Hc\backslash\Hc(A)\}$, the channel coefficients $\Hc(A)$ satisfy the bounded density assumption. Then, 
					\eqm{
							H(A\Tt \mid S,\Hc) \geq H(B\Tt \mid S) + \tau\; o(\log \Pb).
					}
			\end{theorem}
		Note that Theorem~\ref{theo:theorem_ais} also holds if $F(t)$ is replaced with bounded density linear combinations $F^b_a(t)$ (cf. \cite[Remark 5 of Theorem 1]{Davoodi2020_sumset}).   
		Let us now apply  Theorem~\ref{theo:theorem_ais} into our setting.
		The entropy terms are conditioned by the messages $\Wk$ and the caches $\Zk$. 
		Importantly, the caches store only information about the messages $\Wk$ (either coded or not). 
		Hence, $\Zk$ and $\Wk$ are independent of the channel coefficients and satisfy the property of the conditioning random variable $S$ in Theorem~\ref{theo:theorem_ais}.  
		From this fact and~\eqref{eq:y_is_la}-\eqref{eq:y_is_lb}, it follows that 
			\eqm{\label{eq:ss}
				H(\Ybar_k\Tt\mid \Zk,\Wk, \Hc) - H(\Ybar_{k+1}\Tt\mid \Zk,\Wk,\Hc)
						\geq \tau\; o(\log \Pb). 
			}	
		For complex channels, the term $o(\log\Pb)$ becomes $o(\log P)$~\cite{Davoodi2016_TIT_DoF,Davoodi2018_TIT_BC,Davoodi2020_sumset}, which concludes the proof for Lemma~\ref{lem:ais1}. \qed

\end{appendices}

			\bibliographystyle{IEEEtran}												
			\bibliography{IEEEabrv,Literature} 									

\end{document}

%% file: settings_preamble.tex
	\usepackage[T1]{fontenc}		
	\usepackage[utf8]{inputenc}	
	\usepackage[english]{babel}	

	\usepackage{url}
	\usepackage{ifthen}

	\usepackage{amsthm,amssymb}
	\usepackage{mathtools}
	\usepackage{bm}
	\usepackage{nicefrac}
	\usepackage{dsfont} 
	\usepackage[subnum]{cases}  

	\usepackage{subfiles} 

	\usepackage{placeins} 	
	\usepackage{caption}
	\usepackage{subcaption}
	\usepackage{tikz}
		\usetikzlibrary{arrows,positioning,automata}
		\usetikzlibrary{fit,shapes}
		\usetikzlibrary{calc}

	\usepackage{pgfplots}
		\usepgfplotslibrary{fillbetween}
	\usepackage[percent]{overpic}
	\usepackage{pict2e} 
	\usepackage[most]{tcolorbox}
	
	\usepackage{multirow}
	\usepackage{colortbl}
	\usepackage{longtable} 	
	\usepackage{array}
	\usepackage{arydshln}  
	\usepackage{diagbox}
	
	\usepackage[final]{microtype} 
	\usepackage{tablefootnote} 
	\usepackage{enumitem}
	
	\usepackage{color}
	\definecolor{gray}{rgb}{0.6,0.6,0.6}

	\definecolor{blued}{rgb}{0.0,0.5,1}	
	\definecolor{greened}{rgb}{0.28,.71,.46}	
	\usepackage{cite}

	\newtheorem{theorem}{Theorem}
	\newtheorem{lemma}{Lemma}

	\newtheorem{definition}{Definition}
	\newtheorem{remark}{Remark}

	\allowdisplaybreaks  
	\makeatletter 
	\newcommand{\leqnomode}{\tagsleft@true}
	\newcommand{\reqnomode}{\tagsleft@false}
	\makeatother

	\newtoggle{draft}
	\toggletrue{draft}


%% file: Definitions.tex
\definecolor{mycolor1}{rgb}{0.00000,0.44700,0.74100}%
\definecolor{mycolor2}{rgb}{0.85000,0.32500,0.09800}%
\definecolor{mycolor3}{rgb}{0.92900,0.69400,0.12500}%
\definecolor{mycolor4}{rgb}{0.49400,0.18400,0.55600}%
\definecolor{mycolor5}{rgb}{0.300000,0.64314,0.00000}%
\colorlet{mygreen}{green!60!black}%
\colorlet{myblue}{mycolor1!80!black}%

\newcommand{\Ybar}{\ensuremath{\bar{Y}}}

\newcommand{\Xbar}{\ensuremath{\bar{X}}}

\newcommand{\Tt}{^{[\tau]}}

	\definecolor{mycolor1}{rgb}{0.00000,0.44700,0.74100}%
	\definecolor{mycolor2}{rgb}{0.85000,0.32500,0.09800}%
	\definecolor{mycolor3}{rgb}{0.92900,0.69400,0.12500}%
	\definecolor{mycolor4}{rgb}{0.49400,0.18400,0.55600}%
	\definecolor{mycolor5}{rgb}{0.300000,0.64314,0.00000}%
	
\newcommand\itemEq[1][]{%
  \ifx\relax#1\relax  \item \else \item[#1] \fi
  \abovedisplayskip=0pt\abovedisplayshortskip=0pt~\vspace*{-\baselineskip}}
	
\newcommand{\sij}{_{i,j}}
\newcommand{\sji}{_{j,i}}

\newcommand{\myoverset}[2]{\ensuremath{\overset{\mathclap{#1}}{#2}}}

\newcommand{\itb}{\begin{itemize}}
\newcommand{\ite}{\end{itemize}}
\newcommand{\enb}{\begin{enumerate}}
\newcommand{\ene}{\end{enumerate}}
\newcommand{\eqm}[1]{\begin{align}#1\end{align}}

\newcommand{\limpf}{\lim_{P\rightarrow\infty}}
	
\newcommand{\LF}{\left\lfloor}
\newcommand{\RF}{\right\rfloor}

\newcommand{\LB}{\left(}
\newcommand{\RB}{\right)}

\newcommand{\hW}{{{\hat{W}}}}

\newcommand{\Pb}{{\bar{P}}}

\DeclareMathOperator{\Conv}{Conv}

\DeclareMathAlphabet{\mathbit}{OML}{cmr}{bx}{it}
\DeclareMathAlphabet{\mathsf}{OT1}{cmss}{m}{n}
\DeclareMathAlphabet{\mathTXf}{OT1}{cmss}{bx}{it}

\DeclareMathOperator{\Exp}{{\mathbb{E}}}

\DeclareMathOperator{\DoF}{DoF}

\newcommand{\Nb}{{{\mathbb{N}}}}

\newcommand{\Cb}{{{\mathbb{C}}}}
\newcommand{\Ind}{{\mathds{1}}} 

\newcommand{\hv}{\mathbf{h}}

\newcommand{\vv}{\mathbf{v}}

\newcommand{\Bc}{{{\mathcal{B}}}}
\newcommand{\Cc}{{{\mathcal{C}}}}

\newcommand{\Gc}{{{\mathcal{G}}}}
\newcommand{\Hc}{{\mathcal{H}}}

\newcommand{\Kc}{{{\mathcal{K}}}}

\newcommand{\Sc}{{{\mathcal{S}}}}
\newcommand{\Tc}{{{\mathcal{T}}}}

\newcommand{\Wc}{{{\mathcal{W}}}}
\newcommand{\Xc}{{{\mathcal{X}}}}

\newcommand{\Zc}{{{\mathcal{Z}}}}

\newcommand{\bH}{\mathbf{H}}

\newcommand{\bX}{\mathbf{X}}

\newcommand{\bd}{\bm{d}}

\newcommand{\bphi}{\ensuremath{\boldsymbol{\phi}}}

